\documentclass[review]{elsarticle}

\usepackage{hyperref}

\usepackage{color}
\usepackage{newtxtext,newtxmath}
\usepackage{graphicx}
\usepackage{amsmath}
\usepackage{subfigure}
\usepackage{bm}
\usepackage{bbm}
\usepackage{diagbox}

\journal{Carbon}

\begin{document}

\begin{frontmatter}

\title{PAI-graphene: a new topological semimetallic two-dimensional carbon allotrope with highly tunable anisotropic Dirac cones}

\author[uu]{Xin Chen}
\author[ni,uu]{Adrien Bouhon}
\author[hu,au]{Linyang Li\corref{mycorrespondingauthor}}
\ead{linyang.li@hebut.edu.cn}
\author[yu,au]{Fran{\c{c}}ois M. Peeters}
\author[uu]{Biplab Sanyal\corref{mycorrespondingauthor}}
\ead{biplab.sanyal@physics.uu.se}

\cortext[mycorrespondingauthor]{Corresponding author}
\address[uu]{Department of Physics and Astronomy, Uppsala University, Box 516,
751\,20 Uppsala, Sweden}
\address[ni]{Nordic Institute for Theoretical Physics (NORDITA), Stockholm, Sweden}
\address[hu]{School of Science, Hebei University of Technology, Tianjin 300401, People's Republic of China}
\address[au]{Department of Physics, University of Antwerp, Groenenborgerlaan 171, B-2020 Antwerp, Belgium}
\address[yu]{Department of Physics and Astronomy, Key Laboratory of Quantum Information of Yunnan Province, Yunnan University, 650091 Kunming, China}

\begin{abstract}
Using evolutionary algorithm for crystal structure prediction, we present a new stable two-dimensional (2D) carbon allotrope composed of polymerized as-indacenes (PAI) in a zigzag pattern, namely PAI-graphene whose energy is lower than most of the reported 2D allotropes of graphene. Crucially, the crystal structure realizes a nonsymmorphic layer group that enforces a nontrivial global topology of the band structure with two Dirac cones lying perfectly at the Fermi level. The absence of electron/hole pockets makes PAI-graphene a pristine crystalline topological semimetal having anisotropic Fermi velocities with a high value of $7.0 \times 10^{5}$ m/s. We show that while the semimetallic property of the allotrope is robust against the application of strain, the positions of the Dirac cone and the Fermi velocities can be modified significantly with strain. Moreover, by combining strain along both the x- and y-directions, two band inversions take place at $\Gamma$ leading to the annihilation of the Dirac nodes demonstrating the possibility of strain-controlled conversion of a topological semimetal into a semiconductor. Finally we formulate the bulk-boundary correspondence of the topological nodal phase in the form of a generalized Zak-phase argument finding a perfect agreement with the topological edge states computed for different edge-terminations.
\end{abstract}


\end{frontmatter}

\section{Introduction}
Due to the rich diversity of hybridization forms of carbon atoms, many allotropes of carbon exist in zero-dimensional (0D)\cite{318162a0,Kaisereaay1914}, one-dimensional(1D)\cite{354056a0}, two-dimensional (2D)\cite{Novoselov666,Kempkes2019}, and three-dimensional (3D) forms. Graphene as the lowest-energy 2D allotrope of carbon exhibits very interesting unusual physics, e.g., linear dispersion, high carrier mobility, quantum Hall effect, and so on \cite{Novoselov666,nature04235,nature08582,Novoselov1379,nphys245,PhysRevLett.123.036401,KOU2015418}. Inspired by the great success of exfoliating graphene in 2004 \cite{Novoselov666}, many efforts have been conducted to search for other 2D carbon allotropes. Composed of $sp$-$sp^{2}$ hybridized atoms, a series of 2D carbon allotropes, namely graphynes, were proposed\cite{doi:10.1063/1.453405,PhysRevB.58.11009}. These structures exhibit excellent thermal stability and various electronic properties (metallic, semimetallic, and semiconducting)\cite{doi:10.1002/pssb.201046583,PhysRevB.58.11009,PhysRevLett.108.086804,S000862231830900X,MORRESI2020512}. However, due to their high energy, only a few of them have been realized experimentally\cite{10.1039/b922733d,doi:10.1021/jacs.6b12776,Kaisereaay1914}. Therefore, searching for carbon allotropes with low energy is highly demanded.

In addition to graphene and graphynes, many other 2D carbon allotropes have been proposed theoretically by including non-hexagonal rings, such as T-graphene (4-8 rings)\cite{PhysRevLett.108.225505}, penta-graphene (pentagons)\cite{10.1073/pnas.1416591112}, pentaheptites (5-7 rings)\cite{PhysRevB.53.R13303,doi:10.1021/ci000010j}, Haeckelite sheets (5-6-7 rings)\cite{PhysRevLett.84.1716,doi:10.1021/nl049879x}, Hope-graphene (5-6-8 rings), 
ph-graphene \cite{ZHANG2016323} and recently predicted 5-6-7 rings composed phagraphene\cite{doi:10.1021/acs.nanolett.5b02512}, $\psi$-graphene\cite{PhysRevB.75.085432,doi:10.1021/acs.jpclett.7b01364} and SW-graphene\cite{PhysRevB.99.041405}. Some unusual properties have been predicted for these materials, such as negative Poisson's ratio\cite{10.1073/pnas.1416591112,doi:10.1021/acs.jpclett.9b00905}, semimetallic properties\cite{PhysRevLett.108.225505,doi:10.1021/acs.nanolett.5b02512,PhysRevB.99.041405}, and high Li storage capacity\cite{doi:10.1021/acs.jpclett.7b01364}. Most of these allotropes are composed of $sp^{2}$ and $sp^{3}$ hybridized atoms and have lower energy than experimentally synthesized graphynes and are likely to be synthesized in the future.

In computational material discovery, the crystal structure is the most crucial input to get the properties of a material. However, in many conditions, we cannot get the crystal structure easily by experimental techniques. Many efforts have been paid to solve this fundamental problem, and several algorithms have been introduced, such as simulated annealing, metadynamics, genetic algorithms, data mining, etc \cite{Pannetier1990,doi:10.1002/anie.199612861,PhysRevLett.90.075503,C2CE06642D,A901227C,doi:10.1063/1.2210932,JM9930300531,doi:10.1021/acs.jpclett.8b00615,doi:10.1021/jp970984n,PhysRevLett.91.135503}. Among them, evolutionary algorithms are particularly attractive due to their efficiency and reliability. By implementing evolutionary algorithms, many new materials with excellent stability and exciting properties have been predicted \cite{Zhang1502,Oganov2019,PhysRevLett.112.085502,PhysRevB.87.195317,PhysRevMaterials.3.013405,PhysRevLett.110.136403,Zhang2017}.

In this work, we have performed a systematic crystal structure search using the evolutionary algorithm based code  USPEX\cite{doi:10.1063/1.2210932,PhysRevLett.112.085502,PhysRevB.87.195317,GLASS2006713,doi:10.1021/ar1001318,LYAKHOV20131172} for 2D carbon allotropes and found a new 2D carbon allotrope, PAI-graphene. The energy of this allotrope is comparable to graphene and lower than most of the predicted 2D carbon allotropes. We performed phonon calculations and molecular dynamics simulations up to 1500 K and found PAI-graphene to be dynamically stable. Crucially, the crystal structure realizes the nonsymmorphic layer group LG44 (Pbam), which enforces a nontrivial global topology of the band structure with two Dirac cones at the Fermi level. Thanks to the absence of electron/hole pockets, the Dirac nodes lie perfectly at the Fermi level, making PAI-graphene a pristine crystalline topological semimetal. Remarkably, the global band topology is fully controlled by the energy ordering of the irreducible representations (IRREPs) at $\Gamma$, implying that the band topology can be converted through band inversions at $\Gamma$ only. We show that strain induces a band inversion at $\Gamma$ responsible for the transfer of the Dirac points from the high-symmetry line $\overline{\Gamma Y}$ to $\overline{\Gamma X}$ after colliding without annihilation at $\Gamma$. We also show that by applying an additional strain along the x-direction, a second band inversion below the Fermi level takes place at $\Gamma$ inducing the annihilation of the Dirac nodes. We hence prove the strain-controlled conversion of a topological semimetal into a semiconductor. We finally formulate the bulk-boundary correspondence of the topological nodal phase in the form of generalized Zak-phase argument and show the stability of the topological edge states under variations of the edge termination.

\section{Computational details}
In evolutionary structure search, the projector augmented wave (PAW) method\cite{PhysRevB.59.1758,PhysRevB.50.17953} based density functional code VASP\cite{002230939500355X,PhysRevB.54.11169} is used to perform first-principles calculations. The exchange-correlation potential energy was treated within the generalized gradient approximation method in the form of Perdew, Burke, and Ernzerhof (PBE)\cite{PhysRevLett.77.3865}. The wave function was expanded in a plane wave basis with an energy cutoff of 520 eV, and the sampling of the Brillouin zone (BZ) was as dense as $2\pi \times 0.033$ \AA$^{-1}$. Perpendicular to the 2D layer, a vacuum of 20 \AA~was used. To allow for low-buckling structures but maintaining a 2D form, the initial thickness of the region containing the atoms was set to 1 \AA.

When calculating the final low-energy structures, we optimized the structure again and calculated the energetics, and electronic properties using a higher accuracy. The structures were optimized using the conjugate gradient (CG) and RMM-DIIS quasi-Newton algorithms\cite{0009261480803964} until the Hellman-Feynman forces on each atom were less than 0.01 eV/\AA. For all the calculations, BZ sampling denser than $2\pi \times 0.025$ \AA$^{-1}$ was adopted, and the energy convergence parameter was kept as 10$^{-5}$ eV. The phonon calculations were performed using density-functional perturbation theory (DFPT) with the PHONOPY code\cite{phonopy}. The thermal stability was checked using ab initio Born-Oppenheimer molecular dynamics (BOMD) simulations in which the temperature was adjusted by Andersen thermostat\cite{doi:10.1063/1.439486}. To get reliable electronic properties, electronic band structures were calculated with both PBE and the Heyd-Scuseria-Ernzerhof (HSE) hybrid functional within the framework of HSE06\cite{10.1063/1.2404663}. The results of PBE is shown in Figure S4 in the supplementary information (SI).

\section{Results and discussions}
\subsection{Structure and stability}
\begin{figure}
\begin{center}
\includegraphics[scale=1]{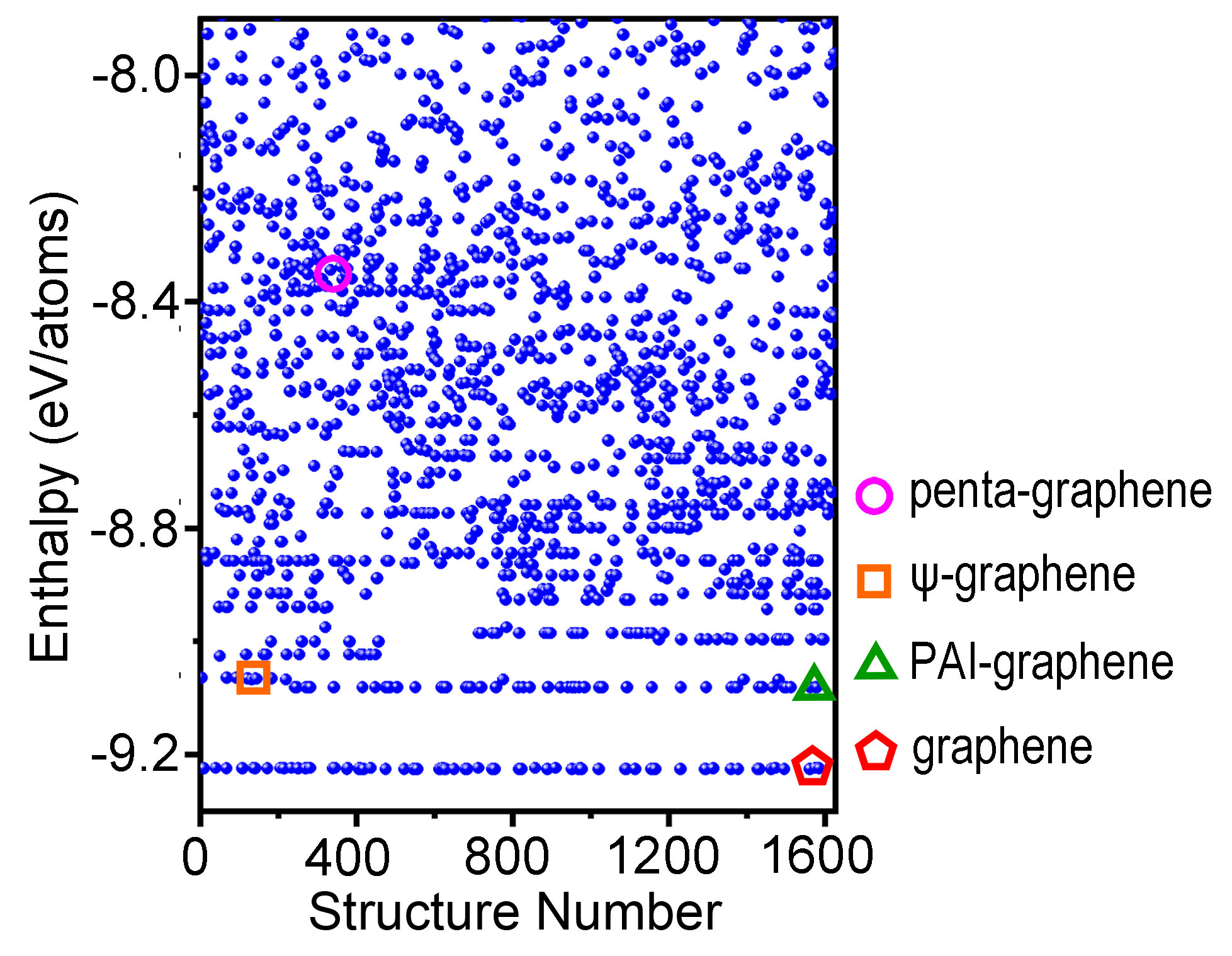}
\caption{The enthalpies of the allotropes found in our evolutionary structure search. The energies of PAI-graphene and several reported structures are indicated by different circles.}
\label{figure1}
\end{center}
\end{figure}

Our evolutionary structure search has reproduced most of the reported 2D carbon allotropes with 2, 3, 4, 6, 8, 12, and 24 atoms/unit cell. The enthalpies of these carbon allotropes are shown in Fig. \ref{figure1}. Among all these structures, graphene exhibits the lowest energy, i.e., -9.227 eV/atom. Though $sp$ and $sp^{3}$ hybridizations exist in many of the produced structures, the lowest-energy structures are usually composed of $sp^{2}$ hybridized atoms. Despite graphene, PAI-graphene with a 24-atom unit cell was found to be the lowest-energy 2D carbon allotrope, with the energy of -9.083 eV/atom, lower than most of the previously proposed planar carbon structures. Compared to the previously reported four most stable 2D carbon allotropes, PAI-graphene is as stable as SW-graphene (-9.085 eV/atom)\cite{PhysRevB.99.041405}, and energetically more favorable than phagraphene (-9.027 eV/atom)\cite{doi:10.1021/acs.nanolett.5b02512}, Octite M1 (-9.029 eV/atom)\cite{Appelhans_2010}, and $\psi$-graphene (-9.069 eV/atom), which were also found in our structure search\cite{doi:10.1021/acs.nanolett.5b02512}.

\begin{figure*}
\begin{center}
\includegraphics[scale=0.7]{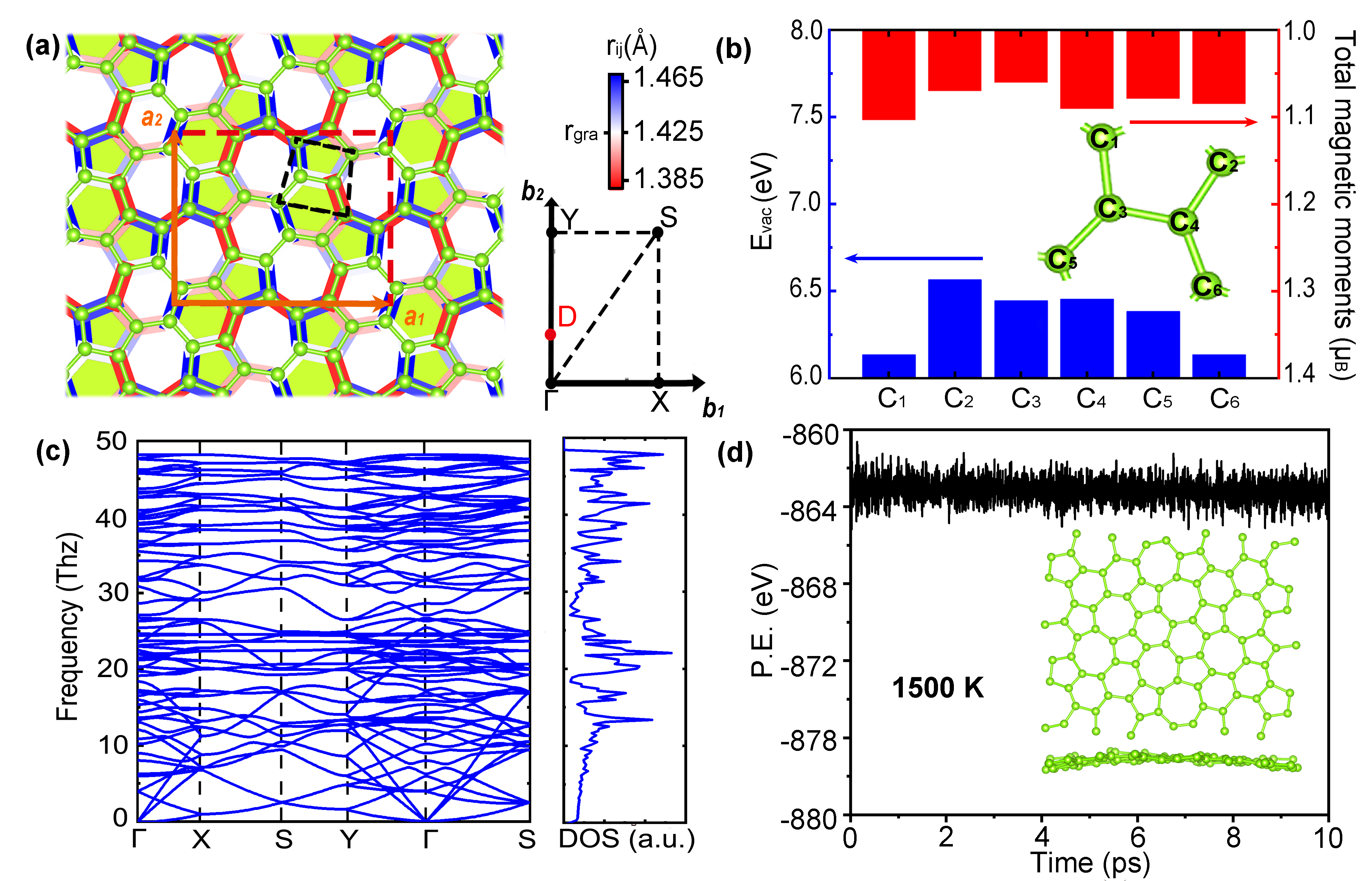}
\caption{(a) The top view of the optimized structure, with its primitive cell marked by red dashed rectangle. {\bf $\vec{a}_{1}$} and {\bf $\vec{a}_{2}$} are the lattice vectors in x- and y-directions, respectively. Green blocks mark the as-indacenes carbon skeleton. The relative bond lengths between atom i and j are illustrated using sticks with different colors, and the value of $r_{ij}$ is given in the color bar, where $r_{gra}$ is the bond length in graphene. The right figure is part of the first BZ with high symmetric points indicated. (b) The vacancy formation energies E$_{vac}$ when removing nonequivalent carbon atoms C$_{i}$, which are enclosed by black dashed lines in (a) and enlarged in the inset with the indication of the C$_{i}$ atoms. (c) Phonon dispersion and phonon density of states (DOS). (d) Potential energy (P.E.) as a function of simulation time t at 1500 K and in the inset, the top and side views of the final structure are presented.}
\label{figure2}
\end{center}
\end{figure*}

The structure of PAI-graphene is shown in Fig. \ref{figure2}(a). PAI-graphene is composed of 5-6-7 carbon rings, similar to the previously predicted four low-energy structures. It can be viewed as a zigzag alignment of the carbon skeletons of as-indacene, a hydrocarbon molecule composed of 5-6-5 carbon rings, with 24 carbon atoms in its primitive cell. PAI-graphene has plane symmetry Pbam (layer group No. 44, i.e., the 2D slice of space group No. 55). It is composed of the two-dimensional primitive orthorhombic Bravais lattice and point group $D_{2h}$ ($mmm$), see the character table (Table S1) in SI. It is nonsymmorphic with two screw axes $\{s_x,s_y\}$ and with, perpendicular to these, two glide planes $\{g_x,g_y\}$, where $s_i = (C_{2i}\vert \boldsymbol{\tau})$ and $g_i = (m_{i}\vert \boldsymbol{\tau})$ with $i=x,y$, and the fractional shift defined in the basis of primitive Bravais vectors as $\boldsymbol{\tau} = (\boldsymbol{a}_1+\boldsymbol{a}_2)/2$ \cite{ITCA,ITCE}. The C-C bond lengths in PAI-graphene are different from those in graphene. The longest and the shortest C-C bonds are 1.468 \AA~ and 1.392 \AA, respectively. There are six non-equivalent atoms, C$_{1}$-C$_{6}$, which are shown in Fig. \ref{figure2}(b). 

Now, we will explore the possibility to induce magnetism in this 2D structure. There is a continuous serach of 2D magnetic materials from evolutionary algorithm\cite{PhysRevB.93.085406,PhysRevB.99.205412}. The other route is to manipulate the structural and electronic properties by introducing defects, e.g., vacancies\cite{RevModPhys.81.109,doi:10.1021/nn1024175, PhysRevB.79.113409,doi:10.1021/acs.jpcc.9b07804}. By removing one carbon atom and optimizing the structures, we investigated the energetics and magnetic properties of PAI-graphene with such a monovacancy. The formation energy of vacancies has been calculated by $E_{\mathrm{vac}}=E_{V}+\frac{1}{2}E_{\mathrm{gra}}-E_{\mathrm{P}}$,
where $E_{V}$ is the total energy of PAI-graphene in a $2 \times 2 \times 1$ supercell geometry with a monovacancy, $E_{\mathrm{gra}}$ is the total energy of a graphene primitive cell, and E$_{\mathrm{P}}$ is the total energy of pristine PAI-graphene in a $2 \times 2 \times 1$ supercell geometry. As shown in Fig. \ref{figure2}(b), removal of C$_{1}$ and C$_{6}$ atoms requires the lowest energy to create a monovacancy, with $E_{V} = 6.13$ eV, much lower than that for graphene, e.g., 6.92 eV \cite{PhysRevB.89.205411}. The total magnetic moments of the six monovacancy defects are not very different, i.e., about 1.1 $\mu_{B}$ per vacancy, smaller than in graphene, i.e., 1.5 $\mu_{B}$ per monovacancy \cite{PhysRevB.89.205411}.

In order to confirm the dynamic stability, phonon dispersion spectrum and phonon DOS were calculated using a $2 \times 2$ supercell, and are shown in Fig. \ref{figure2}(c). There are no modes with imaginary frequencies, and therefore the allotrope is dynamically stable. The longitudinal acoustic (LA) and transverse acoustic (TA) branches are linear near $\Gamma$ point, while the out-of-plane acoustic (ZA) branch shows a quadratic feature without a linear component. This is due to the fact that the XZ and YZ components of the harmonic force constants of PAI-graphene are zero, which is also observed in some other 2D materials\cite{doi:10.1021/acsami.6b04211,doi:10.1021/acs.jpclett.0c00613}. The thermal stability is examined by ab initio BOMD simulations using a $2 \times 2$ supercell with a time step of 1 fs. We followed the system during 10 ps after heating the structure to temperatures of 300 K, 800 K, and 1500 K. As shown in Fig. \ref{figure2}(d) and SI, in the final geometrical framework, the 5-6-7 carbon ring structures are well preserved, and no structural reconstruction occurs in all the three cases. 

Though the $sp^{2}$ graphene allotropes are energetically more favorable than the synthesized carbon materials such as graphynes, fullerenes\cite{318162a0}, and very recently realized C$_{18}$ rings\cite{Kaisereaay1914}, a possible synthesis route is of vital importance for its potential application. Defect topology and molecular assembly were regarded as two possible routes discussed in the literature\cite{doi:10.1021/acs.nanolett.5b02512,doi:10.1021/acs.nanolett.5b02512,PhysRevB.99.041405}. In this work, we propose a synthesis route for PAI-graphene, by polymerizing the carbon skeletons of as-indacenes, i.e, $2 \mathrm{C}_{12} \mathrm{H}_{8} \rightarrow \mathrm{C}_{24}$(PAI-graphene)$ + 8 \mathrm{H}_{2}$. It is an exothermic reaction, in which an energy of 1.70 eV per PAI-graphene unit cell is released.

\subsection{Electronic properties}
\begin{figure*}
\begin{center}
\includegraphics[scale=0.7]{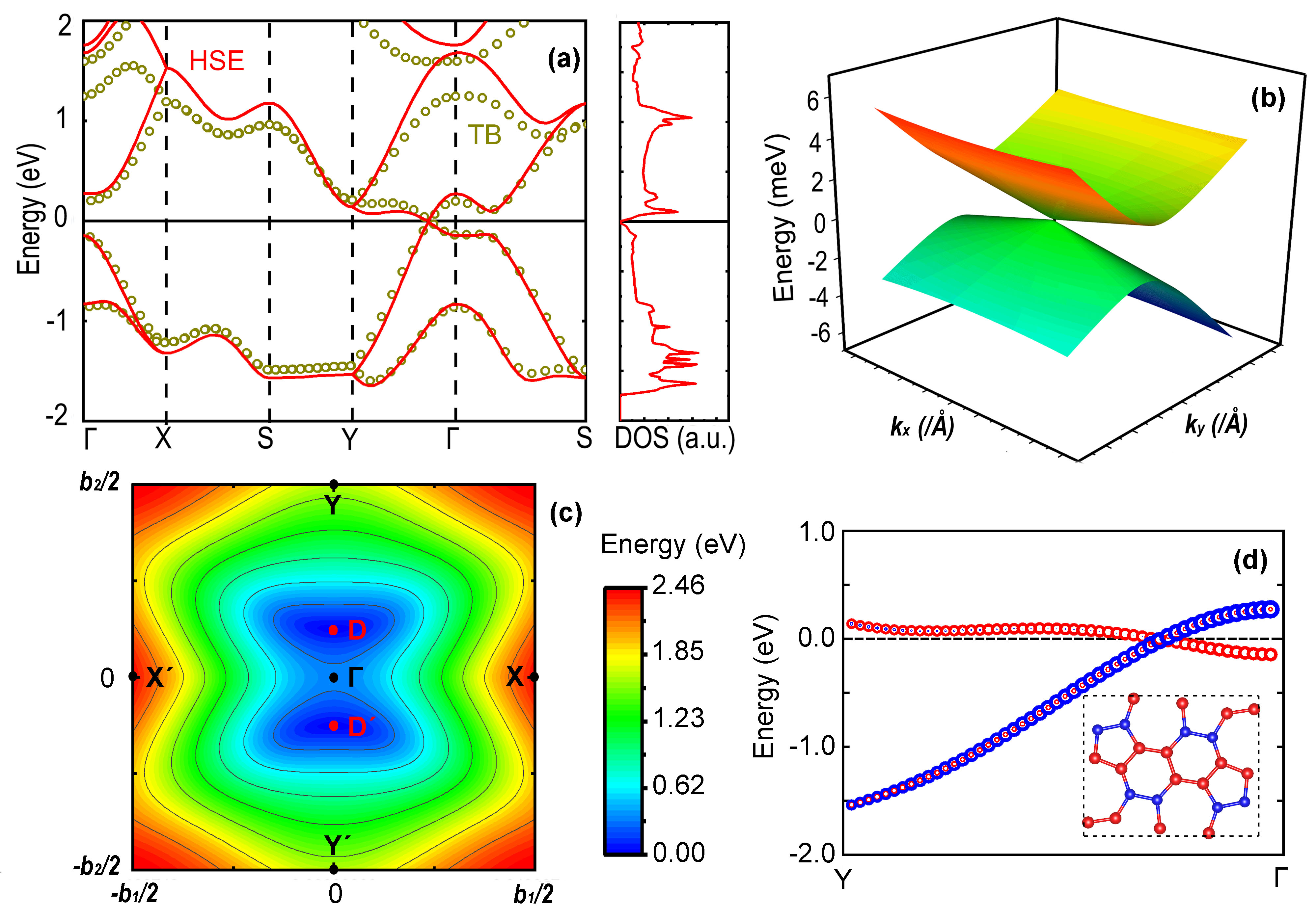}
\caption{(a) The electronic band structures obtained by HSE06 (red lines) and the TB model (brown-green circles). The DOS calculated using HSE06 is shown in the right panel. (b) The 3D band structure in the vicinity of the Dirac cone, as calculated using the TB model. (c) The energy difference between the lowest conduction band and highest valence band, mapped on the first BZ, calculated using the TB model. (d) The atomic orbital projected bands along Y-$\Gamma$. The contributions of the two groups of atoms (in red and in blue) are marked in the corresponding color.}
\label{figure3}
\end{center}
\end{figure*}

The electronic band structure and total electronic density of states (DOS) obtained by HSE06 hybrid functional calculations are shown in Fig. \ref{figure3}. As shown in Fig. \ref{figure3}(a), the valence band maximum (VBM) and conduction band minimum (CBM) of PAI-graphene meet at the Fermi level between the Y and $\Gamma$ points and form a distorted Dirac cone. In the first BZ, there are two Dirac cones, one is between Y and $\Gamma$, and the other one is its centrosymmetric Dirac cone between Y$^{\prime}$ and $\Gamma$ points. The DOS at the Fermi level is zero, which confirms the presence of the distorted Dirac cones and proves that the Dirac cones are not generated from band folding. The Fermi velocities are calculated by linear fitting of the band structures, following $\mathrm{v}_{\mathrm{F}}=\nabla_{|k|} E(k) / \hbar$. The calculated Fermi velocities in the $k_{y}$ direction are $2.5 \times 10^{5}$ m/s (with band slope S$>$0) and $7.0 \times 10^{5}$ m/s (with band slope S$<$0), comparable to that of graphene $1.01 \times 10^{6}$ m/s (HSE) \cite{S000862231830900X}.

To analyze the origin of the distorted Dirac cones, we plotted the atomic orbital projected band structures along the high symmetric path Y-$\Gamma$, in the vicinity of the Fermi level. As shown in Fig. \ref{figure3}(d), the states near the Dirac cones are primarily made up of $p_{z}$ atomic orbitals. The bands with slope S$>$0 are mainly resulting from the carbon atoms shown in red, while the bands with slope S$<$0 are mostly resulting from the carbon atoms in blue. The inversion of the occupied states at the Dirac cones is similar to those in phagraphene and SW-graphene \cite{doi:10.1021/acs.nanolett.5b02512,PhysRevB.99.041405}. Next, we propose a simple tight-binding (TB) model containing only $p_{z}$ atomic orbitals to describe the dispersion of the $\pi$-electrons in the vicinity of the Fermi level. The effective Hamiltonian is taken as Ref. \cite{doi:10.1021/acs.nanolett.5b02512}
\begin{equation}
H_{TB}=\sum _{{\left \langle i,j \right \rangle}} t{_{ij}}c{_{i}^{+}} c{_{j}}+h.c.,
\end{equation}
where $c{_{i}^{+}}$, and $c{_{i}}$ are the creation and annihilation operators of an electron at $p_{z}$ atomic orbitals of the $i^{th}$ carbon atom. $t{_{ij}}$ is the hopping parameter between the electrons at $p_{z}$ orbitals of the nearest neighbors $i^{th}$ atom and $j^{th}$ atom. For graphene and some other carbon structures, limiting ourselves to the nearest neighbor hopping terms give already a good agreement with the DFT band structure. In our case, the valence band and the conduction band are not symmetric around the Fermi level. Therefore, we need to include more interaction terms between carbon atoms, i.e., beyond the nearest neighbor approximation. The parameter $t{_{ij}}$ is distance-dependent, which is determined by $t{_{ij}}=t_{0}exp(q \times (1-d_{ij}/d_{0}))$, where $t_{0} = -2.7$ eV, $q = 2.8$, and $d_{0} = 1.5$ \AA. The parameter $q$ is an adjustment factor. We truncate the terms for the hopping energies less than 0.1 eV. The electronic band structure obtained by diagonalizing the Hamiltonian is shown in Figs. \ref{figure3}(a)-(c). As shown in Fig. \ref{figure3}(a), the TB electronic bandstructure fits the HSE bands well and reproduces the Dirac cones. The electronic energy surface in the vicinity of the Dirac cone is shown in Fig. \ref{figure3}(b). Notably, the highest Fermi velocity is along the $k_{y}$-direction. Moreover, we mapped the energy difference between the highest valence band and the lowest conduction band in the first BZ. As shown in Fig. \ref{figure3}(c), there are two zero points (Dirac cones) in the first BZ, and they are equivalent due to the central symmetry.

\subsection{Global band topology and Dirac nodes}
As it is well known for nonsymmorphic space groups, LG44 hosts degeneracies on the Brillouin zone boundary corresponding to the two-dimensional projective irreducible representations (IRREPs) of the little co-group of the high-symmetry lines $\overline{\text{SX}}$ and $\overline{\text{SY}}$ \cite{BradCrack}, see Fig.~\ref{figure3}{\bf (a)}. These essential degeneracies imply that the bands must be composed of pairs of connected bands, i.e., each pair of bands cannot be separated by an energy gap over the whole Brillouin zone. Furthermore, LG44 has the specificity that these degeneracies have distinct compatibility relations for the IRREPs from $X$ to $\Gamma$ and from $Y$ to $\Gamma$. This allows a higher connectivity of band structures \cite{Zak_EBR3,Zak_EBR4,Zak_2002}, namely additional symmetry protected nodal points may appear inside the Brillouin zone on the high-symmetry line $\overline{\Gamma\text{X}}$ or $\overline{\Gamma\text{Y}}$ such that groups of four bands are connected (i.e., without an energy gap in between). While these nodal points are free to move along the high-symmetry lines, they cannot be removed unless a special ordering of the IRREPs at $\Gamma$ is realized. We indeed show that the existence of these nodal points is completely determined from the ordering in the energy of the IRREPs at $\Gamma$ \cite{Wi2,BBS_nodal_lines}. 

\begin{figure}
\begin{center}
\includegraphics[scale=0.7]{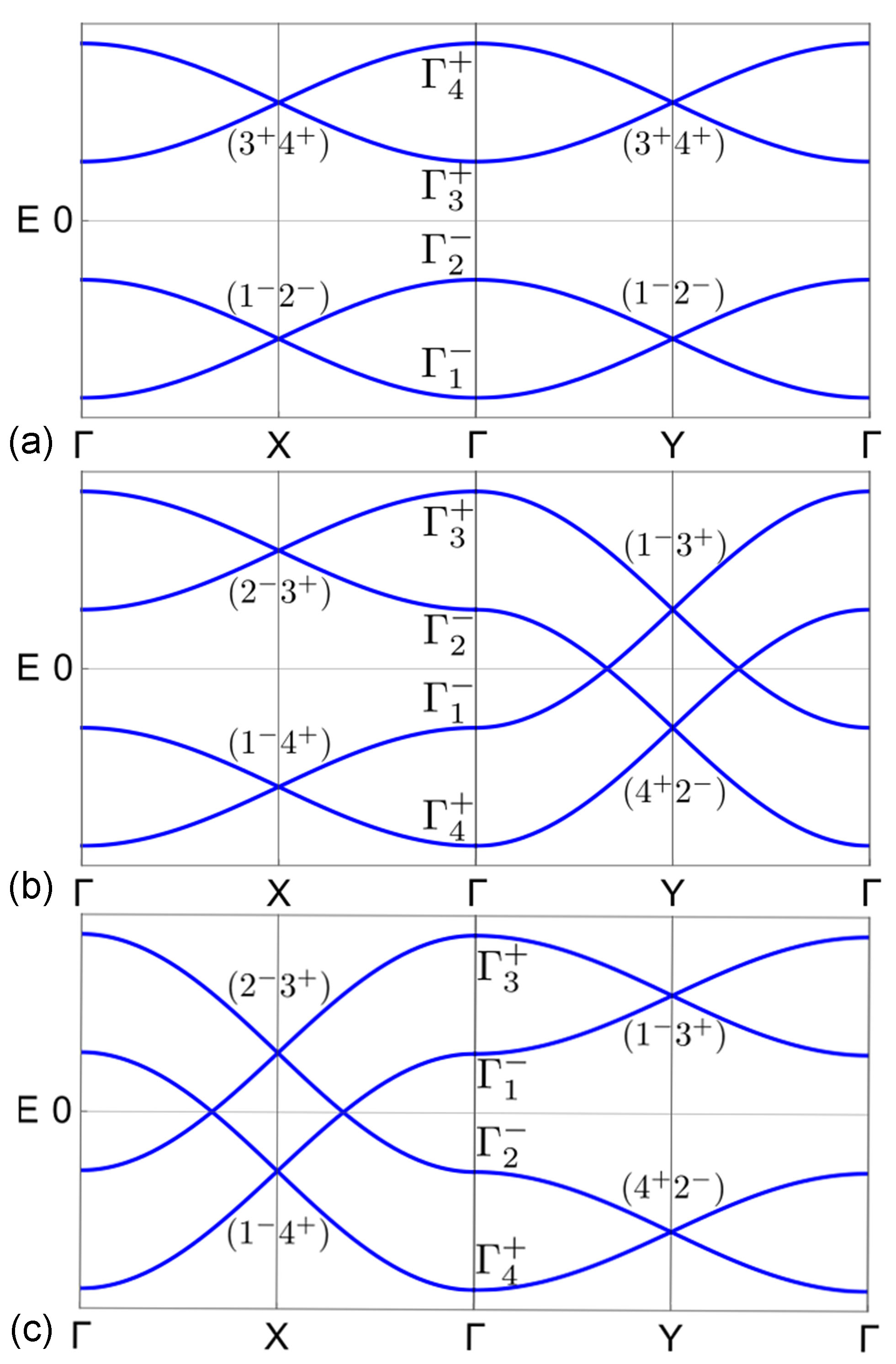}
\caption{Connectivity of band structures along $\overline{\Gamma \text{X}}$ and $\overline{\Gamma \text{Y}}$ determined by the energy ordering of the bands at $\Gamma$. (a) Gapped phase at half-filling with $E^{(1^-)}_1 \leq E^{(2^-)}_2 \leq E^{(3^+)}_3 \leq E^{(4^+)}_4$. (We write $E_i^{(j^{\pm})}$ the energy of the $i$-th band at $\Gamma$ with the IRREP $ \Gamma^{\pm}_j $.) The permutation of bands are $(1^-2^-)$ and $(3^+4^+)$, both along $\overline{\Gamma X}$ and $\overline{\Gamma Y}$. (b) Nodal phase at half-filling with $E^{(4^+)}_1 \leq E^{(1^-)}_2 \leq E^{(2^-)}_3 \leq E^{(3^+)}_4$. The permutation of bands is $(1^-4^+)$ and $(3^+4^+)$ along $\overline{\Gamma X}$, and $(4^+2^-)$ and $(1^-3^+)$ along $\overline{\Gamma Y}$. The nodal points on the high-symmetry line $\overline{\Gamma Y}$ between the bands $1^-$ and $2^-$ are protected by the screw symmetry $s_y$ (equivalently, by the glide symmetry $g_x$). (c) After performing a band inversion between the bands $1^-$ and $2^-$ in (b), the nodal points are moved to the high-symmetry line $\overline{\Gamma X}$.}
\label{figure4}
\end{center}
\end{figure}

The existence of the nodal points inside the Brillouin zone is most efficiently captured in terms of the allowed permutations of any pair of bands across the Brillouin zone \cite{YoungKane_2015,Wi2,BBS_nodal_lines}. Writing $\Gamma_a \in \{\Gamma^{\pm}_j\}_{j=1,2,3,4}$ the (single-valued) IRREP of a band at $\Gamma$ (see the character table Table S1), we write $(\Gamma_a\Gamma_b)$ whenever a $\Gamma_a$-band and a $\Gamma_b$-band at $\Gamma$ are permuted after one shift by a primitive reciprocal lattice vector in the direction $\overline{\Gamma\text{P}}$ ($\text{P}=\text{X},\text{Y}$), i.e., the band labeled $\Gamma_a$ is permuted with the band $\Gamma_b$ as we travel from $\Gamma$ to $\Gamma' = \Gamma + \boldsymbol{b}_{1(2)}$ ($\boldsymbol{b}_{1(2)}$ are the primitive vectors of the reciprocal lattice), see Fig.~\ref{figure4}. Since the $\Gamma_a$-($\Gamma_b$-)band at $\Gamma$ must then be connected to the $\Gamma_b$-($\Gamma_a$-)band at $\Gamma'$, the two branches must cross an odd number of times on $\overline{\Gamma\text{P}}$. These crossings are protected by either of the screw symmetries $s_{x(y)}$ (equivalently the glide symmetries $g_{y(x)}$). 

{\def\arraystretch{1.4}  
\begin{table}[t]
  \begin{tabular}{l | c| c  }
 \hline
 \hline  
  $\overline{\Gamma\text{P}}$ & $\sigma_h$ & $(\Gamma_a,\Gamma_b)$ \\
   \hline  
  $\overline{\Gamma\text{X}}$~ & $+$ &  ~$(1^{+}2^{+})\,,(3^{-}4^{-})\,,(1^{+}4^{-})\,,(2^{+}3^{-})$  \\
  & $-$ &  ~$(1^{-}2^{-})\,,(3^{+}4^{+})\,,(1^{-}4^{+})\,,(2^{-}3^{+})$  \\
   \hline  
  $\overline{\Gamma\text{Y}}$~ & $+$ & ~$(1^{+}2^{+})\,,(3^{-}4^{-})\,,(1^{+}3^{-})\,,(2^{+}4^{-})$  \\
  & $-$ & ~$(1^{-}2^{-})\,,(3^{+}4^{+})\,,(1^{-}3^{+})\,,(2^{-}4^{+})$  \\
 \hline
 \hline
 \end{tabular}
 \caption{\label{permutation_LG44} Band permutation rules along the high-symmetry lines $\{\overline{\Gamma\text{P}}\}_{\text{P=X,Y}}$ expressed in terms of the IRREPs at $\Gamma$ (we use the shortened notation $j^{\pm} \equiv \Gamma^{\pm}_j$, $j=1,2,3,4$).  
}
\end{table}
}

Abbreviating the labeling of the IRREP $\Gamma_j^{\pm}$ as $j^{\pm}$, we label the bands in terms of their IRREPs at $\Gamma$, i.e. the $i$-th band with energy $E_i$ at $\Gamma$ has an IRREP $\Gamma(E_i) = \Gamma^{\pm}_j \in \{1^-,2^-,3^+,4^+\}$. Table \ref{permutation_LG44} lists all the allowed permutations where we have separated them according to their symmetry under basal mirror symmetry $\sigma_h$, see Table S1 in SI. Since all momenta of the 2D Brillouin zone are invariant under $\sigma_h$ (i.e., $\sigma_h \boldsymbol{k}=\boldsymbol{k}$), the mirror eigenvalues are good quantum numbers, and we call the bands either $\sigma_h$-even or $\sigma_h$-odd. In other words, hybridization between $\sigma_h$-even and $\sigma_h$-odd bands is not allowed by symmetry \footnote{This is only true for negligible spin-orbit coupling, as in our case.} so that we can treat each group of bands separately. In the following, we only consider the $\sigma_h$-odd bands since the bands around the Fermi level are entirely formed by the $\pi$-bonds between the $p_z$ orbitals of Carbon.

It is now straightforward to extract the conditions for the existence of nodal points on $\overline{\Gamma\text{X}}$ or $\overline{\Gamma\text{Y}}$. As it is the case in PAI-graphene we assume that a group of four bands around the Fermi level is separated by an energy gap above and below from all the other bands. We label these four bands from below as $E_1(\boldsymbol{k}) \leq E_2(\boldsymbol{k})  \leq E_3(\boldsymbol{k}) \leq E_4(\boldsymbol{k})$. We call the two bands with lower energies ($\{E_1,E_2\}$) the \textit{occupied subspace} and the two bands with higher energies ($\{E_3,E_4\}$) the \textit{unoccupied subspace}. This formal separation actually matches with the physical separation dictated by the Fermi energy (indeed, we have $E_2(\boldsymbol{k})  \leq E_F \leq E_3(\boldsymbol{k})$ over the whole Brillouin zone in the HSE band structure, see Fig.~\ref{figure3}{\bf (a)}).

We have mentioned the essential degeneracies at the Brillouin zone boundaries, i.e., between the bands $E_1$ and $E_2$, and between the bands $E_3$ and $E_4$, forming lines of twofold degeneracy along the high-symmetry lines $\{\overline{\text{SP}}\}_{\text{P=X,Y}}$. In the following, we focus on the crossings between the occupied and the unoccupied subspaces happening at the Fermi level, i.e., the nodal points $\boldsymbol{k}^*$ at which $E_2(\boldsymbol{k}^*) =  E_F = E_3(\boldsymbol{k}^*)$.  

There are then two possible scenario. (i) If the occupied (equivalently, the unoccupied) subspace is composed of either of the following pairs of IRREPs at $\Gamma$, $\{\Gamma_1^- ,\Gamma_2^-\}$ or $\{\Gamma_3^+ ,\Gamma_4^+\}$, Table \ref{permutation_LG44} tells us that the permutations $(1^- 2^-)$ and $(3^+ 4^+)$ are allowed in both directions, $\overline{\Gamma\text{X}}$ and $\overline{\Gamma\text{Y}}$, see the schematic example in Fig.~\ref{figure4}{\bf(a)}. Therefore the occupied subspace can be separated from the unoccupied subspace by a band gap through the whole Brillouin zone, thus corresponding to a semiconducting phase. 

(ii) If the occupied (equivalently, the unoccupied) subspace realizes any other pair of IRREPs at $\Gamma$, i.e., among $\{\Gamma_1^- ,\Gamma_3^+\}$, $\{\Gamma_1^- ,\Gamma_4^+\}$, $\{\Gamma_2^- ,\Gamma_3^+\}$, or $\{\Gamma_2^- ,\Gamma_4^+\}$, Table \ref{permutation_LG44} tells us that the permutation along $\overline{\Gamma\text{X}}$ within the occupied (equivalently, unoccupied) subspace cannot match with the permutation along $\overline{\Gamma\text{Y}}$. It then follows that the occupied subspace cannot be disconnected from the unoccupied subspace along both directions, and there must be nodal points between the bands $E_2$ and $E_3$ along one direction. This thus corresponds to a symmetry protected semi-metallic phase. 

We illustrate the later scenario with an example. Let us assume that the two-band occupied subspace is composed of the bands $\Gamma(E_1) = 4^+$ and $\Gamma(E_2) = 1^-$, see Fig.~\ref{figure4}{\bf(b)}. Table \ref{permutation_LG44} tells us that the permutation $(1^- 4^+)$ is allowed along $\overline{\Gamma\text{X}}$, while it is not allowed along $\overline{\Gamma\text{Y}}$. Indeed, along $\overline{\Gamma\text{Y}}$ the bands $\{1^-,4^+\}$ can only be permuted with the bands $\{2^-,3^+\}$. It then follows that the tow-band unoccupied subspace must be composed of the pair $\{2^-,3^+\}$. Taking $\Gamma(E_3) = 2^-$ and $\Gamma(E_4) = 3^+$, we obtain the permutations $(4^+2^-)$ and $(1^- 3^+)$ along $\overline{\Gamma\text{Y}}$. Thus the branch $E^{(1^-)}_2(\boldsymbol{k})$ must cross the branch $E^{(2^-)}_3(\boldsymbol{k})$ along $\overline{\Gamma\text{Y}}$, see Fig.~\ref{figure4}{\bf(b)} (we write $E_i^{(j^{\pm})}$ the branch with $\Gamma(E_i) = \Gamma_j^{\pm}$). This crossing is protected by symmetry since the Bloch eigenstates of the $1^-$-branch are even under $s_y$ ($g_x$) and the Bloch eigenstates of the $2^-$-branch are odd under $s_y$ ($g_x$). This example precisely explains the stable crossing observed at the Fermi level in Fig.~\ref{figure3}{\bf (a)}.

\subsection{Strain-induced electronic phase transition}
\begin{figure}
\begin{center}
\includegraphics[scale=0.7]{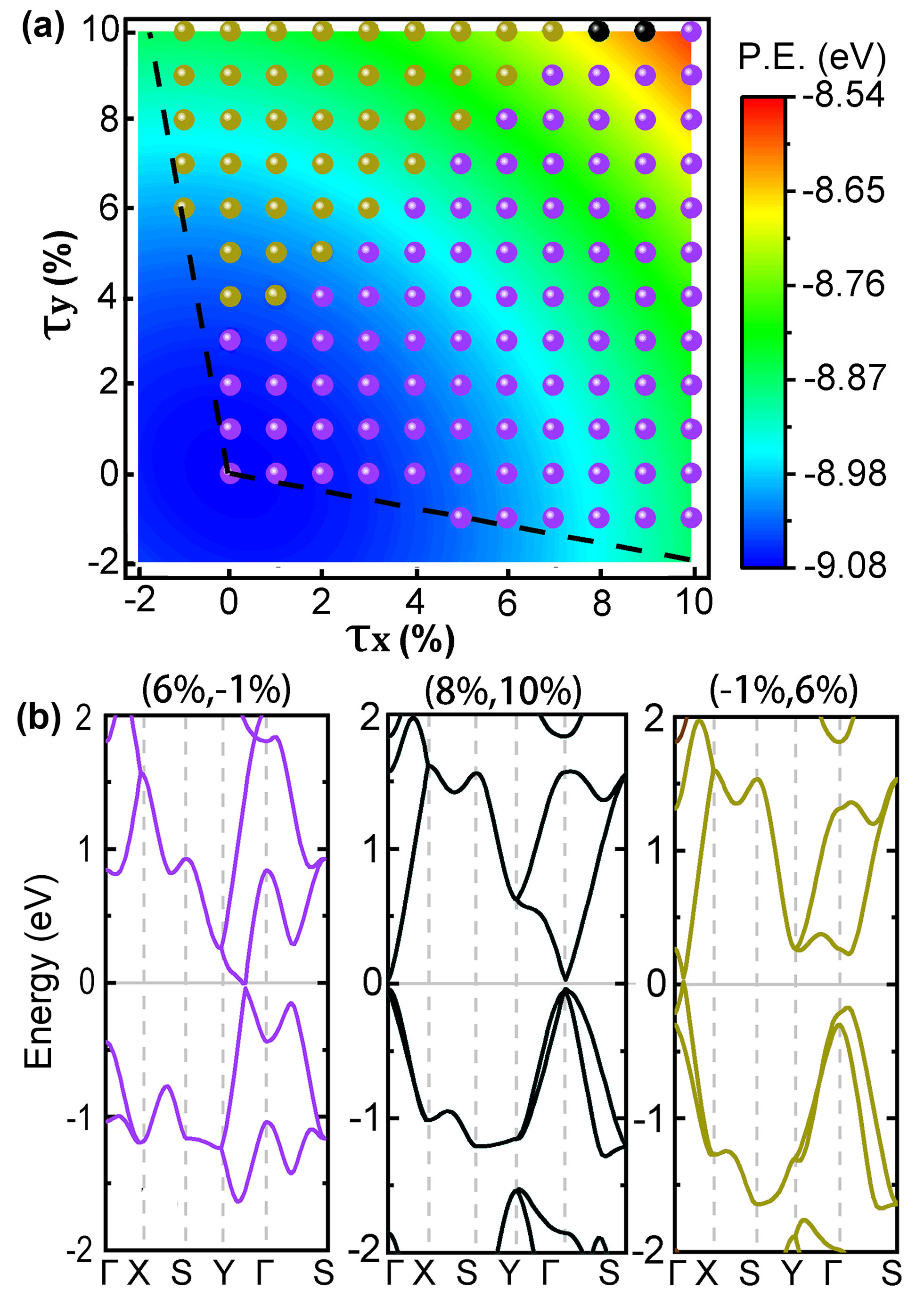}
\caption{(a) Contour plot of the potential energy (P.E.) of PAI-graphene under biaxial strain. The structures with Dirac cones along $Y-\Gamma$ and along $X-\Gamma$ are indicated by purple and brown points. The semiconducting phases are indicated by black points. (b) The band structure of PAI-graphene under strain $\left(\tau_{x}, \tau_{y}\right)=(6 \%,-1 \%)$, $\left(\tau_{x}, \tau_{y}\right)=(8 \%, 10 \%)$, and $\left(\tau_{x}, \tau_{y}\right)=(-1 \%, 6 \%)$ illustrating the three different electronic band structures.}
\label{figure5}
\end{center}
\end{figure}

The effect of external stain in x- and y-direction on the electronic properties of PAI-graphene is systematically investigated by changing the lattice vector from ($\mathbf{a}_{1}$, $\mathbf{a}_{2}$) to ((1+$\tau_{x}$)$\mathbf{a}_{1}$,  (1+$\tau_{y}$)$\mathbf{a}_{2}$). We mapped the potential energies (P.E.) of PAI-graphene under biaxial strain in the range $-2 \% \leq \tau_{x}, \tau_{y} \leq 10 \%$, as shown in Fig. \ref{figure5}(a), and found that the energy changes continuously. By applying axial tensile strain $\tau_{x}$ ($\tau_{y}$) in x (y) direction, the structure will generate a compressive strain $\tau_{y}$ ($\tau_{x}$) in y (x) direction which can be determined from the tangent of the contour line in y (x)-direction. Those tangent points are on the two black dashed lines in Fig. \ref{figure5}(a). In addition, more detailed mechanical properties of PAI-graphene are investigated and discussed in SI.

Though it is possible to apply a small external compressive strain to 2D materials, applying external tensile strain is much easier to be achieved\cite{srep16108,srep08441}. By applying external tensile strain ($\tau_{x}$, $\tau_{y}$), three different regimes for electronic band structures can be found, as indicated in Fig. \ref{figure5}(a), and their corresponding band structures are shown in Fig. \ref{figure5}(b). We have observed that the semi-metallic property is conserved in the purple and brown regimes, with two Dirac points moving under strain. The two Dirac points get very close near the line $\tau_{y}=0.830\tau_{x}+0.202$. Below this line we have the situation as for pristine PAI-graphene as illustrated for $\left(\tau_{x}, \tau_{y}\right)=(6 \%,-1 \%)$, where the Dirac cones are situated on $Y -\Gamma$ and $Y^{\prime}-\Gamma$, marked by purple points in Fig. \ref{figure5}(a). Above the line, the electronic structure is illustrated for $\left(\tau_{x}, \tau_{y}\right)=(-1 \%, 6 \%)$, where the Dirac cones sit on $X-\Gamma$ and $X^{\prime}-\Gamma$, marked by brown points in Fig. \ref{figure5}(a). Moreover, under a strain of $8 \% \leq \tau_{x} \leq 9\%$ and $\tau_{y} \approx 10 \%$, as shown by the black dots in Fig. \ref{figure5}(a), PAI-graphene can be tuned into a direct gap semiconductor.

The obviously different electronic properties induced by external strain sometimes comes from a strain-induced structural phase transition \cite{PhysRevLett.91.135501}. To check whether a structural phase transition has happened under strain, we further computed the phonon spectrum of PAI-graphene under the strain of $\left(\tau_{x}, \tau_{y}\right)=(6\%, -1\%)$, $\left(\tau_{x}, \tau_{y}\right)=(8\%, 10\%)$, and $\left(\tau_{x}, \tau_{y}\right)=(-1\%, 6\%)$, corresponding to the three different electronic phases. As shown in Figure S5 in SI, there is no imaginary frequency found in these phonon spectra, confirming their dynamical stability, and showing that the different electronic properties are not origin from structural phase transition.

The Dirac cones on $X-\Gamma$ show high Fermi velocities. With varying tensile strain, the Fermi velocity can be significantly tuned. At $\left(\tau_{x}, \tau_{y}\right)=(-1 \%, 10 \%)$, the Fermi velocity is increased to $9.7 \times 10^{5}$ m/s. Further calculations show that at $\left(\tau_{x}, \tau_{y}\right)=(-2 \%, 12 \%)$, the Fermi velocity can reach up to $1.05 \times 10^{6}$ m/s, comparable to that of graphene \cite{LIN2017816}.

\begin{figure}
\begin{center}
\includegraphics[scale=0.7]{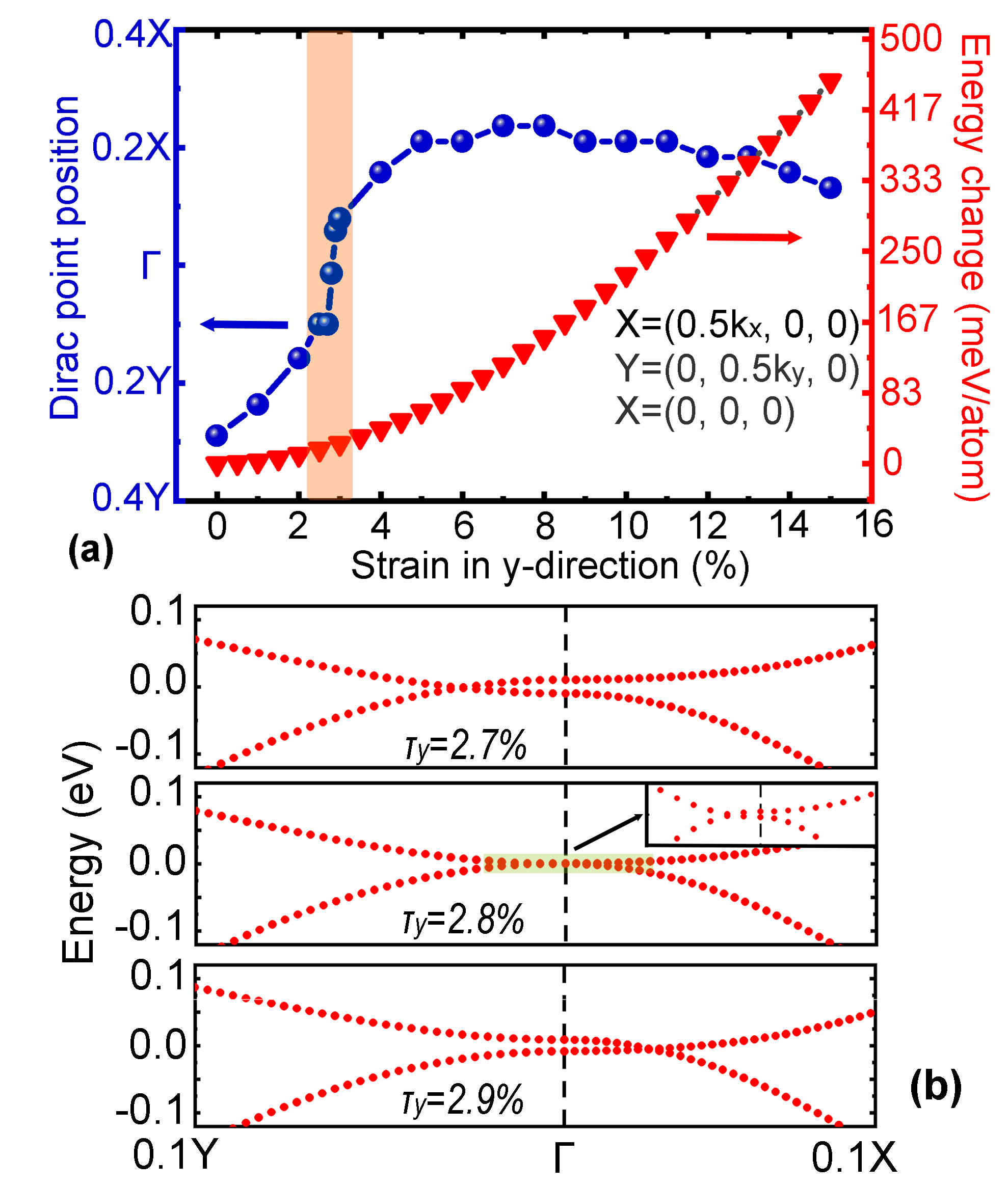}
\caption{(a) The position of the Dirac point (blue) and the energy change (red) under different y-direction uniaxial strain. (b) The detailed band structure near $\Gamma$ point under uniaxial strain $\tau_{y}=2.7\%, 2.8\%$, and $2.9\%$}.
\label{figure6}
\end{center}
\end{figure}

To see the movement of the Dirac cones in detail, we apply uniaxial strain along y-direction (fully relaxed in the x-direction). As shown in Fig. \ref{figure6}, from $\tau_y = 0 \%$ to $\tau_y = 7 \%$, the Dirac point moves from 0.35Y to $\Gamma$, and then from $\Gamma$ to 0.22X, and from $\tau_y = 7 \%$ to $\tau_y = 15 \%$, the Dirac point moves back to 0.18X. Notably, near $\tau_y = 2.8 \%$, the Dirac points are very close to $\Gamma$ point. In this area, the change in Dirac point's position (dark green line in Fig. \ref{figure6}(a)) is very steep, even a $0.1 \%$ strain (energy changes by 1.4 meV/atom) can switch the Dirac points between $X-\Gamma$ and $Y-\Gamma$, as shown in Fig. \ref{figure6}(b). This novel feature makes PAI-graphene a promising material as a sensor with high sensitivity .

We explain the origin of the trajectory of the Dirac nodes under strain from the algebraic rules of band permutations that were exposed above. Under strain, we have seen that the bands $E_2$ and $E_3$ are inverted at $\Gamma$, i.e., the IRREPs ordering is changed as 

\begin{equation}
\begin{aligned}
E_1^{(4^+)} \leq E_2^{(1^-)} \leq &E_3^{(2^-)} \leq E_4^{(3^+)} \longrightarrow \\
&E_1^{(4^+)} \leq E_2^{(2^-)} \leq E_3^{(1^-)} \leq E_4^{(3^+)}\;.
\end{aligned}
\end{equation}
As a consequence, following Table \ref{permutation_LG44}, the occupied subspace must be permuted with the occupied subspace along $\overline{\Gamma\text{X}}$, leading to nodal points between $E_2^{(2^-)}$ and $E_3^{(1^-)}$ on $\overline{\Gamma\text{X}}$. This is illustrated by the schematic examples in Fig.~\ref{figure4}{\bf(b)} and {\bf(c)}, i.e., under the band inversion through the Fermi level at $\Gamma$ the Dirac points move from $\overline{\Gamma Y}$ {\bf(b)} to $\overline{\Gamma X}$ {\bf(c)}. This is also confirmed by the DFT results under strain shown in Fig.~\ref{figure5}{\bf (b)}. 

We conclude that the existence of the Dirac points inside the Brillouin zone is dictated by the ordering in the energy of the IRREPs at $\Gamma$. Also, any perturbation of the system that conserves all the symmetries of LG44 and does not change the IRREPs ordering at $\Gamma$ cannot change the global band topology of the system, i.e., either a semiconductor or a topological semimetal.

\subsection{Bulk-boundary correspondence}

As the previous discussion shows, the Dirac points at the Fermi level are not accidental. It is also well known that the Dirac points in graphene are characterized by a $\pi$-Berry phase, i.e., the Berry phase computed over a loop encircling a single nodal point. The combined symmetry $C_{2z}\mathcal{T}$, i.e.~$\pi$ rotation along the $z$-axis combined with time reversal, (or $I\mathcal{T}$, i.e.~inversion combined with time reversal) requires that the Berry phase factor $e^{i \gamma_B[l]}$ be real for any base loop $l$, there is thus a $\mathbb{Z}_2$ quantization of the Berry phase, i.e., $\gamma_B[l] \,(\mathrm{mod}\,2\pi) \in \{0 ,\pi\}$ \cite{bouhon2020geometric}. As long as $C_{2z}\mathcal{T}$ (or $I\mathcal{T}$) symmetry is preserved, any base loop characterized by a $\pi$-Berry phase must enclose one Dirac point. We conclude that the $\pi$-Berry phase is a topological invariant that characterizes the stability of one Dirac node in the band structure of the bulk material.  

\begin{figure}
\begin{center}
\includegraphics[scale=0.7]{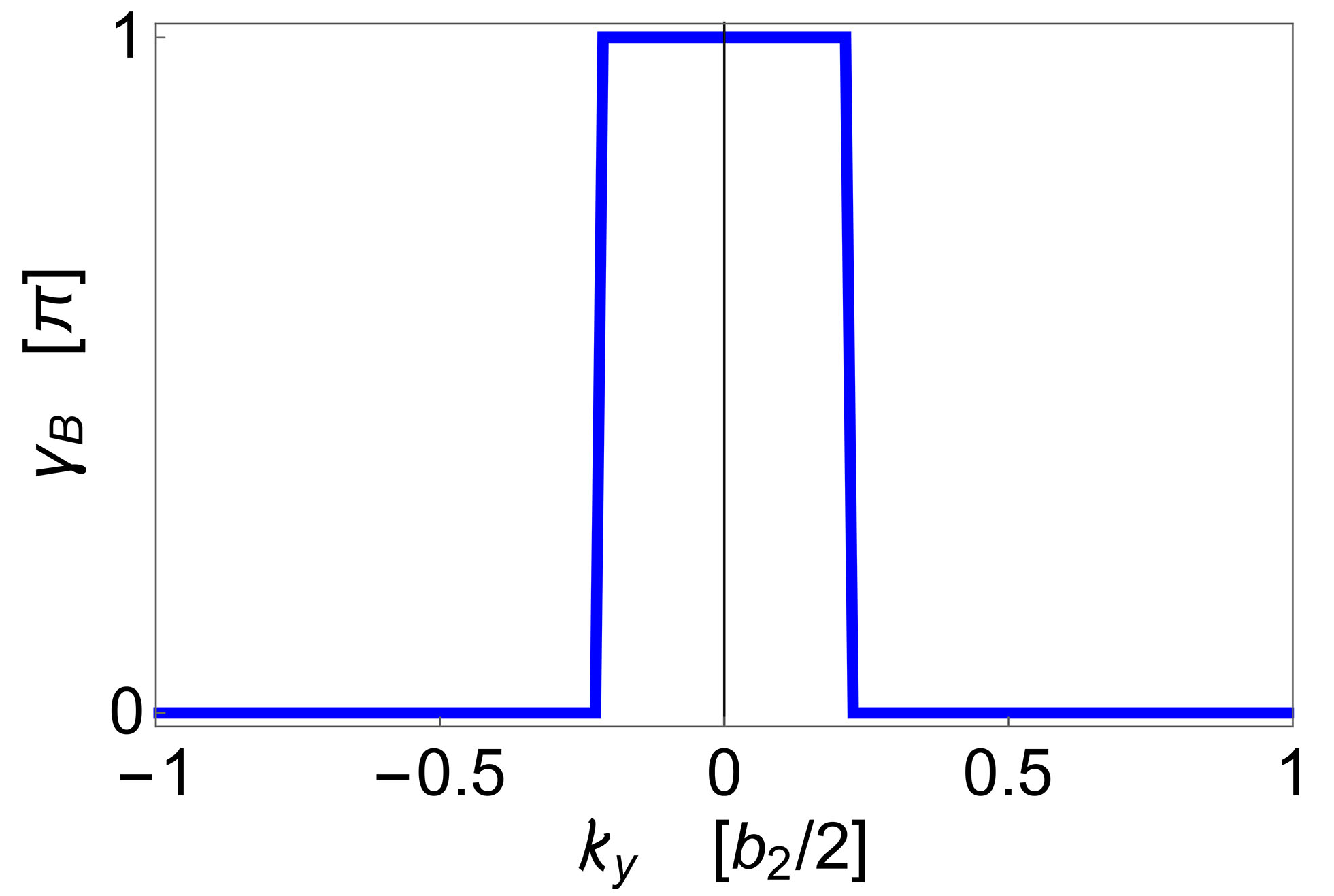}
\caption{$\mathbb{Z}_2$ quantized Berry phase over $l_{k_y} = \bigcup\limits_{k_x\in [-b_1/2,b_1/2]} (k_x,k_y)$ for $k_y\in  [-b_2/2,b_2/2]$ (with $b_i =\vert \boldsymbol{b}_i\vert$, $i,1,2$). Cutting the system along the $y$-axis, i.e.~perpendicular to the $x$-axis, a $\pi$-Berry phase at $k_y$ indicates the existence of one subgap edge state at $k_y$ in the edge band structure by the virtue of the bulk-boundary correspondence principle.}
\label{figure7}
\end{center}
\end{figure}

\begin{figure*}
\begin{center}
\includegraphics[scale=0.7]{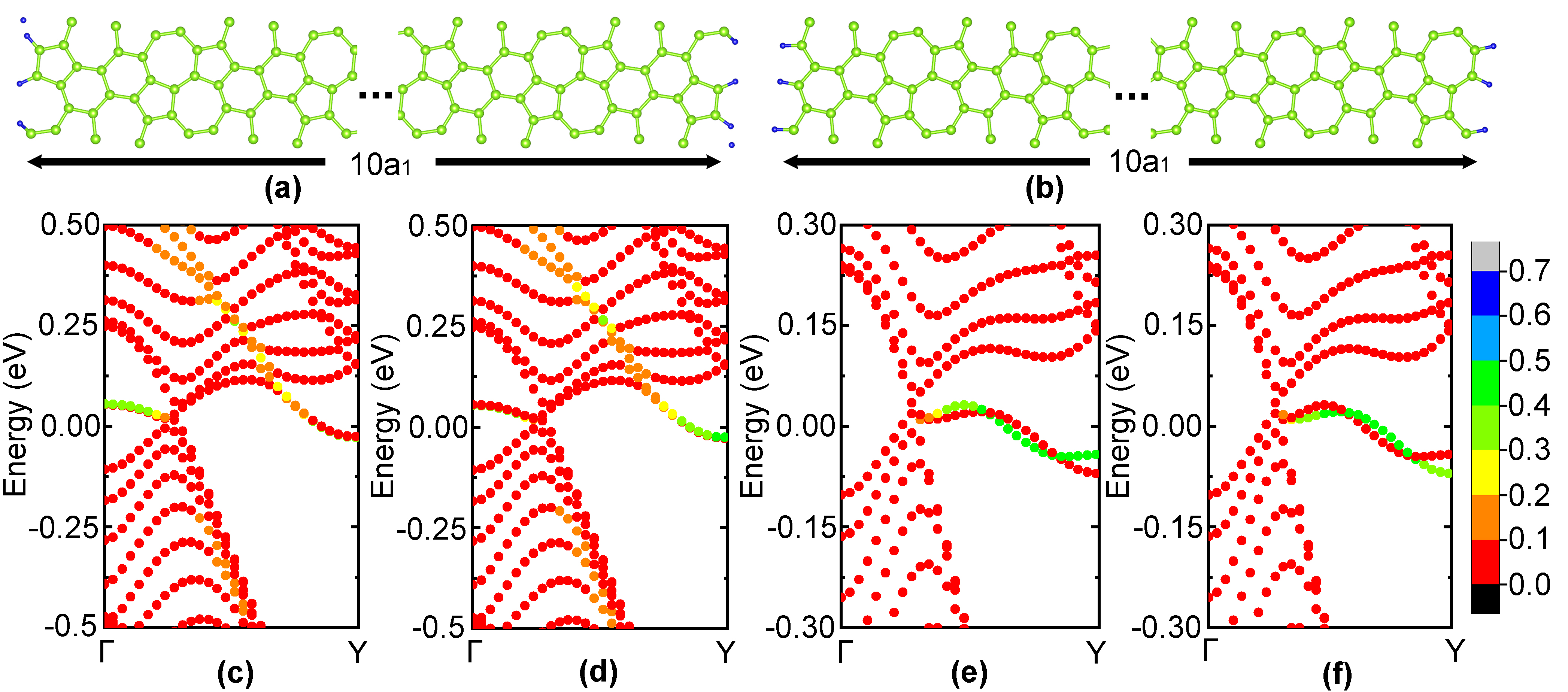}
\caption{(a) and (b) are the structures of nanoribbon with symmetric and asymmetric edges. (c) and (d) show the projected bands of left and right edges of the nanoribbon with symmetric edges, respectively. (e) and (f) present the projected bands of left and right edges of the nanoribbon with asymmetric edges, respectively.}
\label{figure8}
\end{center}
\end{figure*}


From this characterization of the bulk topology, we can predict the existence of topological edge states by invoking the bulk-boundary correspondence \cite{BBC_avila,BBC_graf_porta}.  

First, we reformulate the bulk topological invariant so that it is adapted for the prediction of the topological edge states in a given geometry. Let us consider non-contractible base loops crossing the Brillouin zone from one point of $\overline{\text{SX}}$ to one point of $\overline{\text{SX}}+\boldsymbol{b}_1$ perpendicularly to $\overline{\Gamma\text{Y}}$, i.e., $l_{k_y} = \bigcup\limits_{k_x\in [-b_1/2,b_1/2]} (k_x,k_y)$ for $k_y\in  [-b_2/2,b_2/2]$, with $b_i =\vert \boldsymbol{b}_i\vert$, $i,1,2$. We now compute the Berry phase over these base loops, i.e., $\gamma_B[l_{k_y}]$, as we scan $k_y$ through $[-b_2/2,b_2/2]$. Since PAI-graphene has $C_{2z}\mathcal{T}$ symmetry, the Berry phase is $\mathbb{Z}_2$ quantized. Whenever the base loop $l_{k_y}$ crosses one Dirac node, the Berry phase must jump by $\pi$. This is also known as the Zak phase \cite{PhysRevLett.62.2747}. We show the computation of the Berry phase (Zak phase) as  function of $k_y$ in Fig.~\ref{figure7}. 


In the following, we assume that the choice of origin for the unit cells is a center of $C_{2z}$ symmetry. There are two possible choices corresponding to two distinct Wyckoff positions: (i) the middle of the bonds connecting two pentagons (WP1), (ii) the middle of the bonds shared by two hexagons (WP2). In the following, we choose WP1 as the unit cell origin. We know from Ref.~\cite{PhysRevLett.62.2747} that the Berry phases give the center of charge of the occupied bands. A $0$-Berry phase over a loop $l_{k_y}$ indicates that Wannier centers of the occupied electrons are centered at WP1. A $\pi$-Berry phase over a loop $l_{k_y}$ indicates that the Wannier centers of the occupied electrons are shifted by one half of the unit cell in the $x$-direction \cite{PhysRevLett.62.2747}, i.e., centered at WP2. 

Cutting the system in a ribbon geometry with the two edges parallel to $\boldsymbol{a}_2$, $k_y$ is still a good quantum number of the edge spectrum. The bulk-boundary correspondence \cite{PhysRevLett.62.2747} tells that whenever $\gamma_B[l_{k_y}] = \pi$, there must be one subgap edge state in the edge spectrum at $k_y$. This is a manifestation of the mismatch between the number of charges of the system and the number of available sites at which the occupied charges are localized. Note that the number of charges of the system corresponds to the number of lattice sites contained in the ribbon along the finite $x$-direction since each site contributes by one $p_z$ orbital. 

We show the edge spectrum computed from DFT in Fig.~\ref{figure8} for two distinct ribbon geometries. Considering the huge numerical efforts in the HSE method, we use stretching to change the PBE energy band of the material into a semimetal, and then calculate the edge state of the nanoribbon. The bulk band spectrum of the stretched structure is shown in Figure S4 (b). Fig.~\ref{figure8}{\bf (a)} shows a section of the ribbon where the two edges cross WP1, i.e., they cut a $C_{2z}$-symmetric bond. While there are as many WP1 sites as there are occupied charges centered at WP1, i.e., with $0$-Berry phase, there is a mismatch for the occupied charges centered at WP2, i.e., with $\pi$-Berry phase. From the bulk Berry phase Fig.~\ref{figure7} we then predict that there must be one subgap edge state for every $k_y$ with a $\pi$-Berry phase. The edge spectrum at both edges, Fig.~\ref{figure8}{\bf (c)-(d)}, directly confirms this (note the subgap branches reaching $Y$ that are not topological and can be removed by changing the hydrogenation of the edges). 

We now show how a change of the edge termination can affect the spectral structure of the topological edge states. Let us consider the ribbon geometry of Fig.~\ref{figure8}{\bf (b)}, where the edges are now cut away from WP1. This changes the counting of the available sites, and incidentally, it flips the bulk-boundary correspondence: it is now a $0$-Berry phase that indicates the existence of a topological subspace state. This is readily confirmed by the computed edge spectrum at both edges, Fig.~\ref{figure8}{\bf (e)-(f)}.      

Importantly, and contrary to the ancillary subgap branches of Fig.~\ref{figure8}{\bf (c)-(d)} that can be removed adiabatically, the topological edge branches are stable as long as the $C_{2z}\mathcal{T}$ symmetry of the system is preserved. This concludes our discussion of the direct manifestation of the nontrivial topology of the Dirac nodes in PAI-graphene.

\section{Conclusions}
In summary, using evolutionary crystal structure search and first-principles calculations, we found a new planar carbon allotrope, namely PAI-graphene (Polymerized As-Indacenes). The energy of PAI-graphene is lower than most of the reported 2D carbon allotropes. Phonon calculations and molecular dynamics simulations confirm its dynamical and thermal stability. Our HSE06 hybrid functional calculations indicate that PAI-graphene is a semimetal with distorted Dirac cones. A TB model is constructed to describe these Dirac cones and is able to provide physical insight on the origin of these Dirac cones.

Making use of representation theory for the non-symmorphic layer group LG44, we have derived the nontrivial global band topology of the system. We have shown that the energy ordering of the IRREPs at $\Gamma$ directly indicates the existence of two Dirac cones at the Fermi level, making PAI-graphene a perfect example of a topological semimetal.

It is also observed that external biaxial tensile strain preserves the semimetallic property. However, the position and the anisotropic properties of the Dirac cones are significantly modified under strain. By using tensile strain in the y-direction, the Fermi velocities can reach high values, which may work as a favorable condition for applications in high-speed electronic devices. Moreover, PAI-graphene has large and anisotropic Young's modulus, which promises useful applications in nanodevices.

As a manifestation of the global band topology, the trajectory of the Dirac cones under strain can be readily understood as the effect of a band inversion through the Fermi level at $\Gamma$. Similarly, the disappearance of the Dirac cones at higher stain can be readily understood as resulting from a double band inversion at $\Gamma$. This demonstrates the strain-controlled conversion of a topological semimetal into a semiconductor. 

Furthermore, we formulated the bulk-boundary correspondence in the form of a generalized Zak-phase argument. We showed that the Dirac cones must be accompanied by topological subgap edge states. Also, by studying two distinct ribbon geometries, we revealed how the spectral structures of the topological edge states depend on the edge termination. 

\section*{Acknowledgement}
We thank S. Nahas,  for helpful discussions. This work is supported by the project grant (2016-05366) and Swedish Research Links program grant (2017-05447) from the Swedish Research Council, the Fonds voor Wetenschappelijk Onderzoek (FWO-Vl) and the FLAG-ERA project TRANS 2D TMD. X.C. thanks China scholarship council for financial support (No. 201606220031). X.C. and B.S. acknowledge SNIC-UPPMAX, SNIC-HPC2N, and SNIC-NSC centers under the Swedish National Infrastructure for Computing (SNIC) resources for the allocation of time in high-performance supercomputers. Moreover, supercomputing resources from PRACE DECI-15 project DYNAMAT are gratefully acknowledged. 

\section*{Appendix A. Supplementary data}
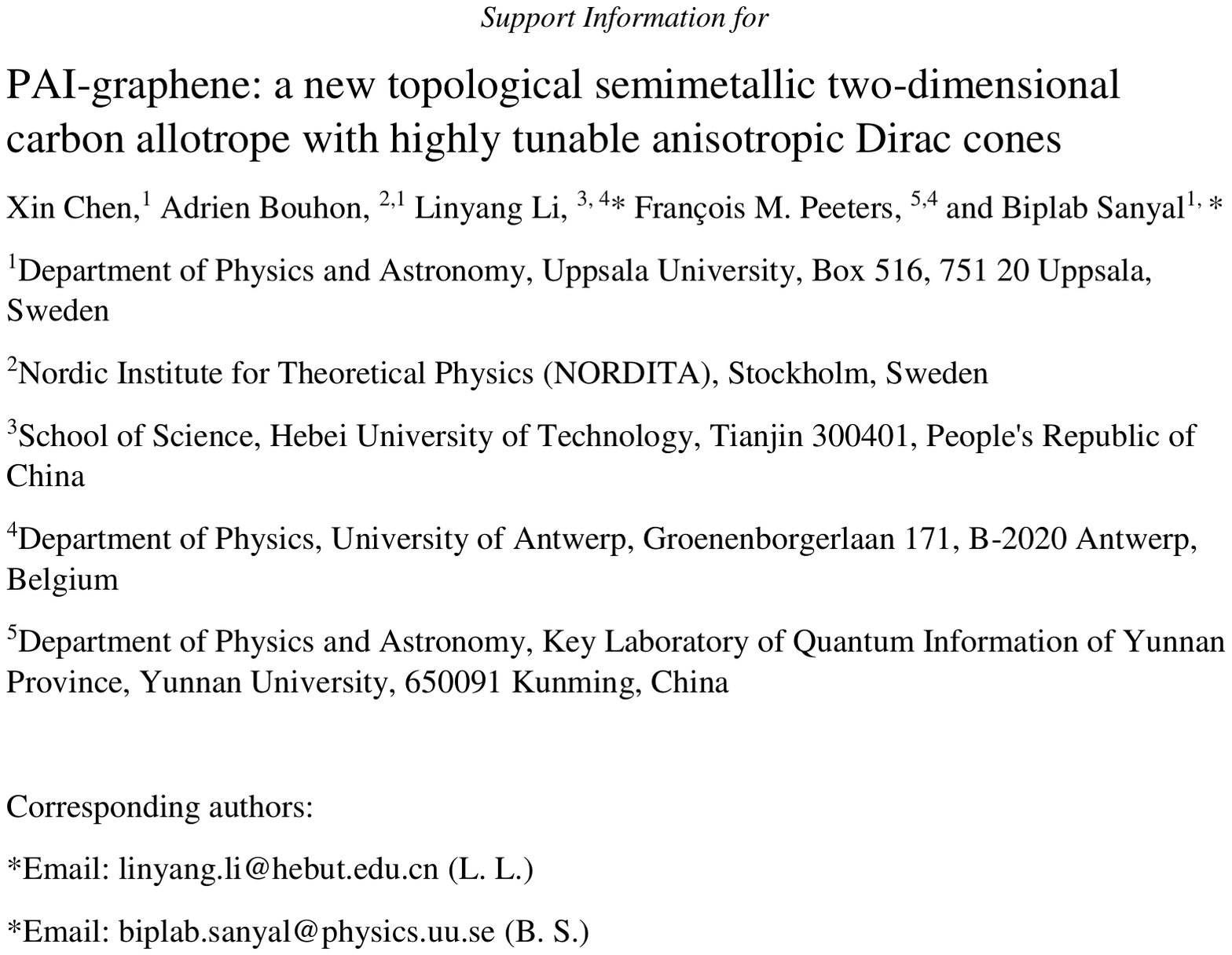: There are six parts: Part I is the result of the BOMD simulations at 300 K and 800 K. Part II contains the details of the strain-induced insulating properties. Part III gives information on the position of the atoms in PAI-graphene. Part IV describes the mechanical properties, including the Young's modulus and Poisson's ratio of PAI-graphene. Part V is the electronic band structure and DOS by PBE calculations. Part VI is the character table of the point group $D_{2h}$. Part VII is the phonon spectra of PAI-graphene under strain.

\bibliography{main}

\end{document}